# Quantum phase transition and destruction of Kondo effect in pressurized $SmB_6$

Yazhou Zhou[1], Qi Wu[1], Priscila F. S. Rosa[2,3], Rong Yu[4,8], Jing Guo[1], Wei Yi[1], Shan Zhang[1], Zhe Wang[1], Honghong Wang[1], Shu Cai[1], Ke Yang[5], Aiguo Li[5], Zheng Jiang[5], Shuo Zhang[5], Xiangjun Wei[5], Yuying Huang[5], Yi-feng Yang[1,7], Zachary Fisk[2], Qimiao Si[1,6], Liling Sun†[1,7] & Zhongxian Zhao[1,7]

[1]*Institute of Physics and Beijing National Laboratory for Condensed Matter Physics, Chinese Academy of Sciences, Beijing 100190, China*
[2]*Department of Physics and Astronomy, University of California, Irvine, CA92697, USA*
[3]*Los Alamos National Laboratory, Los Alamos, New Mexico 87545, USA*
[4]*Department of Physics, Renmin University of China, Beijing 100872, China*
[5]*Shanghai Synchrotron Radiation Facilities, Shanghai Institute of Applied Physics, Chinese Academy of Sciences, Shanghai 201204, China*
[6]*Department of Physics & Astronomy, Rice University, Houston, Texas 77005, USA*
[7]*Collaborative Innovation Center of Quantum Matter, Beijing 100190, China*
[8]*Department of Physics and Astronomy, Shanghai Jiao Tong University, Shanghai 200240, China and Collaborative Innovation Center of Advanced Microstructures, Nanjing 210093, China*

$SmB_6$ is a candidate material that promises to elucidate the connection between strong correlations and topological electronic states, which is a major challenge in condensed matter physics. The electron correlations are responsible for the development of multiple gaps in $SmB_6$, whose understanding is sorely needed. Here we do so by studying the evolutions of the gaps and related properties under pressure. Our measurements of the valence, Hall effect and electrical resistivity clearly identify the gap which is associated with the bulk Kondo hybridization and, moreover, uncover a pressure-induced quantum phase transition from a putative topological Kondo insulating state to a Fermi-liquid state at ~4 GPa. We provide the evidence for the transition by a large change in the Hall number, a diverging tendency of the electron-electron scattering coefficient and, thereby, a destruction of the Kondo entanglement in the ground state. These effects take place in a mixed-valence background. Our results raise the new prospect for studying topological electronic states in quantum critical materials settings.

Samarium hexaboride ($SmB_6$) is a paradigm Kondo insulator (KI) with a simple cubic structure, comprising a $B_6$ octahedral framework and Sm ions[1]. It has in common with other KIs that an insulating gap opens upon cooling due to the hybridization between the localized *f*-electrons and conduction electrons[2-4], but differs from the other KIs by the presence of a low temperature resistance plateau which presented a puzzle for decades[5,6]. Recent theoretical studies suggest that the resistance plateau may be associated with the existence of an exotic metallic surface state, and that $SmB_6$ could be a new class of topological insulator with strong electron correlations, namely a topological Kondo insulator (TKI)[7,8]. A metallic surface state indeed has been confirmed by a variety of measurements[9-16], but whether it is protected by a non-trivial topology remains to be established[17]. Surprisingly, various energy gaps ranging from 5-15 meV have been observed from ambient-pressure resistivity, angle resolved photoemission spectroscopy (ARPES), and scanning tunneling spectroscopy (STS) experiments[6,18-22], it is still unclear whether some of these gaps are generated by the bulk Kondo hybridization between 4*f* electrons and conduction electrons, or are associated with the surface state. The newly discovered quantum oscillation phenomena, which could be related to an unusual Fermi surface in $SmB_6$ at low temperature, yield an additional puzzle[23,24]. In order to make progress, it is crucially important to understand the nature of the multiple electronic gaps.

An important means to understand Kondo hybridization gap is the tuning of Kondo insulators through a quantum phase transition[4], in particular through the phase diagram proposed for such systems at zero temperature[25,26]. For studying quantum phase transition, it is known that high pressure is a 'clean' way of continuously tuning the crystal and electronic structures without introducing additional chemical complexity[27,28]. Although high-pressure phenomena of $SmB_6$ had been extensively investigated before the proposal of topological conducting surface state[5,6,20,29-32], new experiments under high pressure are needed to elucidate the instability of the metallic surface state and Kondo hybridization state. In this study we use our advanced high-pressure facilities, which include the capacity of resistance measurements down to 0.3 K, to carry out such investigations on the mixed-valent $SmB_6$.

We first performed *in-situ* high pressure resistance measurements on single crystal $SmB_6$. As demonstrated in Figs.1a and 1b, the resistivity of the sample A and sample B show the same behavior at pressures below 4 GPa, *i.e.* the resistivity increases continuously upon decreasing temperature and then displays a resistivity plateau below 3K, which is consistent with the results reported previously[5,6,20]. Upon further increasing pressure above 4 GPa, the resistivity shows a remarkable drop at low temperature, which manifests a pressure-induced insulator-metal transition. Interestingly, the onset temperature ($T^*$) of the resistance plateau exhibits a downward trend with increasing pressure, and disappears at ~ 4 GPa (Fig.1c). Above 4 GPa, while the resistivity vs. temperature becomes metallic-like at low temperatures

(T<~20K), it still exhibits an insulating-like behavior at higher temperatures (T>~20K) for pressures up to ~10 GPa, above which it crosses over to metallic behavior (Fig.1a and Fig.S1, Supplementary Information). In the pressure range investigated, no crystal structure phase transition is observed (Fig.S2, Supplementary Information), indicating that the pressure-induced resistance evolution stems from electron-electron interactions.

To determine the nature of the ground state above 4GPa, measurements at low temperatures are necessary. Our newly developed cryostat, with the capability of performing high pressure studies down to 0.3 K, allows us to clarify this issue. By fitting the resistivity-temperature (*R-T*) data from 0.3 K to 4 K with the power law, $\rho(T)=\rho_0+AT^n$ (where $\rho_0$ is the residual resistivity, $A$ the electron scattering coefficient and $n$ the exponent), we find that below 0.7 K the *R-T* curve at 4 GPa follows a Fermi liquid (FL) behavior with $n=2$ (Fig.2a). A linear *R-T* behavior at 4 GPa is only found in the temperature range of 1.5K - 4K (Fig.2a), in agreement with prior measurements at higher temperatures (>1.5K)[20]. The value of the exponent *n* measured in the range of 0.7 K -1.5 K lies between 1 and 2. Moreover, fits to the *R-T* data measured down to 0.3 K for pressures above 4 GPa, to as high as 19.6 GPa, were also made (Fig. 2b, 2c and Fig.S3, Supplementary Information). The FL temperatures ($T_{FL}$) at different pressures are determined by the highest temperature for a $\rho(T)=\rho_0+AT^2$ fitting. The *A* coefficient as a function of pressure is shown in Fig.2d. A strong tendency towards divergence is found when approaching the threshold pressure (~4 GPa) for the insulator-metal quantum phase transition.

The pressure-temperature phase diagram from our transport data is shown in Fig.3a. Three regions appear in this phase diagram. The region below 4 GPa represents the putative topological Kondo insulating (P-TKI) state featuring the resistance plateau. In the intermediate-pressure region between 4 GPa and 10 GPa, the low temperature state is a FL, but the higher temperature state appears non-metallic. In the region above 10 GPa, the sample is metallic up to room temperature but is still in a FL state at low temperature, where long-range magnetic ordered state is observed[29,30]. In order to clarify the response of the system to pressure, we made fits to our resistivity data by the Arrhenius equation (Fig.S4, Supplementary Information), and extrapolated the energy gap at each pressure value. In line with the multiplicity of the gaps, we identify two energy gaps $E_i$ (small gap) and $E_h$ (large gap), when performing the fit of the data below and above ~20 K, respectively. As shown in the inset of Fig.3a, their ambient-pressure values are closely consistent with the reported results determined by resistance measurements[6,20,21]. It is seen that both $E_h$ and $E_i$ decrease monotonically upon increasing pressure. At ~4 GPa where the insulator-metal transition occurs (Fig.1), the $E_i$ gap vanishes and the resistance plateau disappears. By contrast, the $E_h$ gap still exists up to ~10 GPa.

We now consider the nature of the states in the three regions of the phase diagram and the connections among them. For SmB$_6$, the unusual phenomena occur in a mixed valence state with $f$ electron configuration between *4f$^6$* and *4f$^5$+5d*. And this mixed-valance state varies with pressure in a coordinated way, *i.e.* the concentration of conduction electrons (*5d*) and/or localized magnetic ion (*4f$^5$*) increase or decrease simultaneously with the change of the mean valence. To understand the pressure-induced resistance behavior, the corresponding evolution of the two energy gaps and the ground state of the phases in the intermediate-pressure and high-pressure regions, we performed *in-situ* high pressure synchrotron x-ray absorption and Hall coefficient measurements. Representative $L_{III}$-edge spectra collected at different pressures are presented in Fig.S5 (Supplementary Information). The mean valance ($v$) as a function of pressure is plotted in Fig.3b. It can be seen that $v$ increases monotonously from ambient pressure to ~10 GPa. Particularly across 4GPa, the mean $v$ shows an increase from 2.55 to 2.62. The pressure-induced valence change leads us to suggest that an appropriate mixed valence state is crucial for the low temperature behavior in $SmB_6$.

To address the nature of the quantum phase transition in this mixed valent background, we show the results of our Hall measurements at a fixed temperature of 1.7 K for an extended pressure range to ~20 GPa (Fig.3b). The inverse Hall coefficient undergoes a large change across ~4GPa. Comparing this low-temperature isothermal behavior with its counterparts at higher temperatures demonstrates that the feature sharpens as the temperature is lowered (Fig. 3b, inset). Even though the Hall number is expected to contain contributions from both surface and bulk carriers, our results are compatible with a rapid change of the carrier number in the bulk at~4 GPa, which indicates a Kondo destruction[4,26]. Thus, the vanishing $E_i$ reflects the (asymptotic low-temperature) hybridization gap associated with the Kondo-singlet formation in the ground state.

The nature of the quantum phase transition will also be manifested in the behavior at pressures above ~4GPa. We have therefore analyzed the pressure dependence of the $T^2$ coefficient of the electrical resistivity, *i.e.* the *A* coefficient. At the highest pressure measured, 19.6 GPa, the *A* coefficient is equal to 0.0046 $\mu\Omega\cdot$cm/K$^2$. It increases as pressure is lowered, reaching 3.3 $\mu\Omega\cdot$cm/K$^2$ at 4 GPa. This dramatic enhancement, by a factor of about 720, is shown in Fig. 2d. Following the Kadowaki-Woods relation[33], this corresponds to an increase of the effective carrier mass, $m^*$, by a factor of about 27. By analogy with the behavior of heavy fermion metals[4,27], this strong tendency of divergence of the effective mass as the pressure approaches the threshold value for the transition provides compelling evidence for the destruction of the Kondo effect at 4 GPa.

Our results lead to the overall phase diagram shown in Fig. 3a. As pressure is increased towards 4GPa, the Kondo hybridization gap $E_i$ decreases to zero. The resistance plateau, which signifies the metallic surface state, only exists in this pressure range. The suppression of the Kondo effect when the pressure approaches 4GPa from below is accompanied by the divergence tendency of the effective mass as it approaches 4GPa from above. At the same time, the gap $E_h$ remains finite across 4GPa, and only gradually decreases upon further increasing the pressure towards ~10 GPa, where it smoothly crosses over to the metallic-like Kondo-scattering behavior at high temperatures.-In the high pressure region above 10 GPa, the increase of the mean valence, *i.e.* the increase of magnetic Sm$^{3+}$ ions, promotes the development of long-ranged magnetic order, which is consistent with the previous nuclear forward scattering and specific heat measurements[29,30]. The corresponding dependence of the magnetic transition temperature, $T_M$, as a function of pressure, taken from Refs. [29,30], are also shown in the phase diagram in Fig. 3a; how this transition line changes as the pressure is decreased is an important issue that needs to be clarified by future experiments.

In summary, we have studied the exotic high pressure behavior of $SmB_6$ through comprehensive *in-situ* high pressure measurements of resistance, Hall coefficient and synchrotron x-ray diffraction and absorption. We find a pressure-induced quantum phase transition from the ambient-pressure state of a possibly topological Kondo insulator to a Fermi-liquid state at ~4 GPa, which occurs in the background of mixed valency. Our results provide evidence for the destruction of the Kondo effect at this quantum phase transition. Because such an effect coincides with the collapse of the resistance plateau, our findings point towards an exciting prospect for studying topological electronic states in quantum critical materials settings.

**Methods**

High quality single crystals of $SmB_6$ were grown by the Al flux method, as described in Ref. [11]. Pressure was generated by a diamond anvil cell with two opposing anvils sitting on the Be-Cu supporting plates. Diamond anvils with 300 μm flat and non-magnetic rhenium gaskets with 100 μm diameter hole were employed for different runs of the high-pressure studies. The standard four-probe method was applied on the (001) facet of single crystal $SmB_6$ for all high pressure transport measurements. To keep the sample in a quasi-hydrostatic pressure environment, NaCl powder was employed as a pressure transmitting medium for the high-pressure resistance and Hall coefficient effect measurements.

High pressure x-ray diffraction (XRD) and x-ray absorption spectroscopy (XAS) experiments were performed at beamline 4W2 at the Beijing Synchrotron Radiation Facility and at beamline 14W1 at the Shanghai Synchrotron Radiation Facility, respectively. Diamonds with low birefringence were selected for the experiments. A monochromatic X-ray beam with a wavelength of 0.6199 Å was adopted for all XRD measurements. To maintain the sample in a hydrostatic pressure environment, silicon oil was used as pressure transmitting medium in the high-pressure XRD and XAS measurements. Pressure was determined by the ruby fluorescence method[34].

**Acknowledgements**

We thank J. D. Thompson, V. A. Sidorov, M. Aronson and J. H. Pixley for helpful discussions. The work in China was supported by the NSF of China (Grants No. 91321207, No. 11427805, No. 11404384, No. 11522435) and the Strategic Priority Research Program (B) of the Chinese Academy of Sciences (Grants No. XDB07020300, No. XDB07020200). R.Y. was supported by the NSF of China (Grant No. 11374361), and the Fundamental Research Funds for the Central Universities and the Research Funds of Renmin University of China (Grant No. 14XNLF08). Work at Los Alamos was performed under the auspices of the U.S. Department of Energy, Division of Materials Sciences and Engineering. P. F. S. Rosa acknowledges the support of a Director's Postdoctoral Fellowship that is funded by the Los Alamos LDRD program and the FAPESP Grant 2013/2018-0. Work at Rice University was supported by the ARO Grant No. W911NF-14-1-0525 and the Robert A. Welch Foundation Grant No. C-1411.

**Author Information**

† Correspondence and requests for materials should be addressed to L.S. (llsun@iphy.ac.cn)

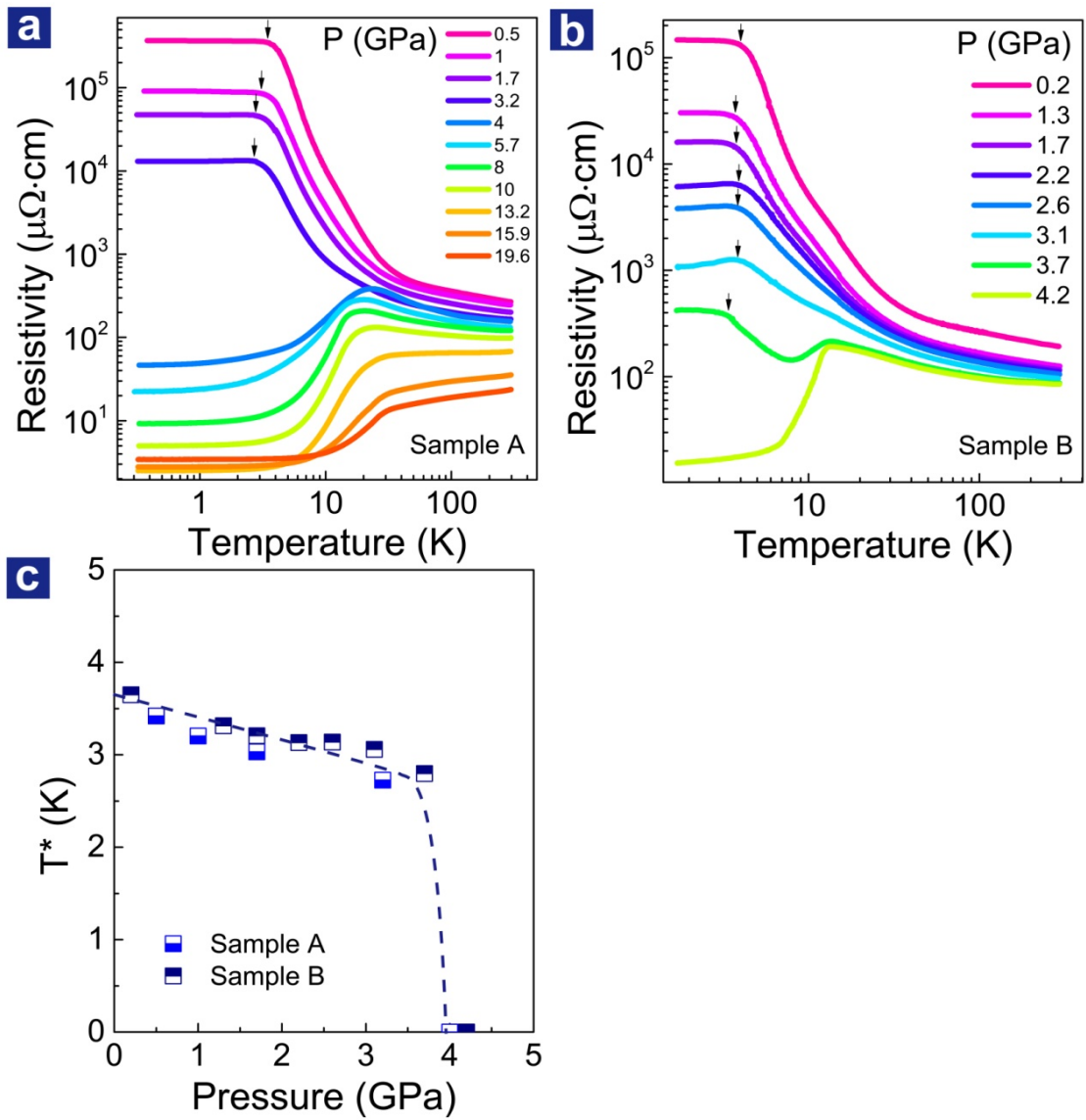

**Figure 1 Resistivity-temperature data and onset temperature of resistance plateau ($T^*$) obtained at different pressures**. (a) Temperature dependence of electrical resistivity in the sample A for pressures ranging from 0.5 GPa to 19.6 GPa. (b) Resistivity as a function of temperature in the sample B for pressures ranging from 0.2 GPa to 4.2 GPa. The arrows in (a) and (b) specify $T^*$, the onset temperature for the resistance plateau. (c) Plot of $T^*$ versus pressure extracted from the resistance measurements.

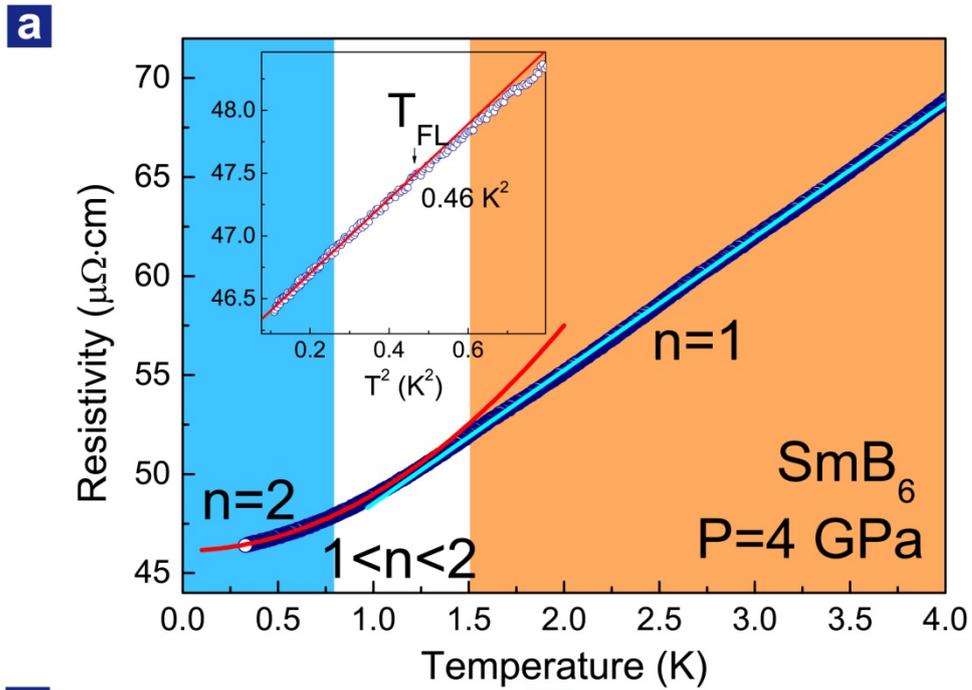
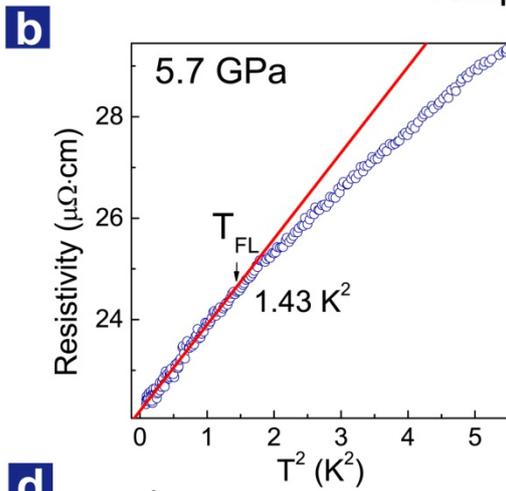
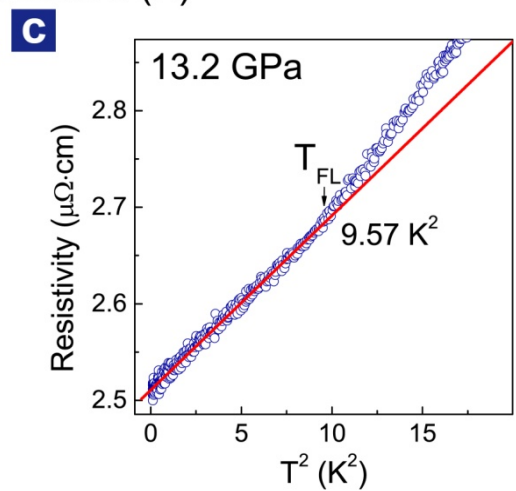
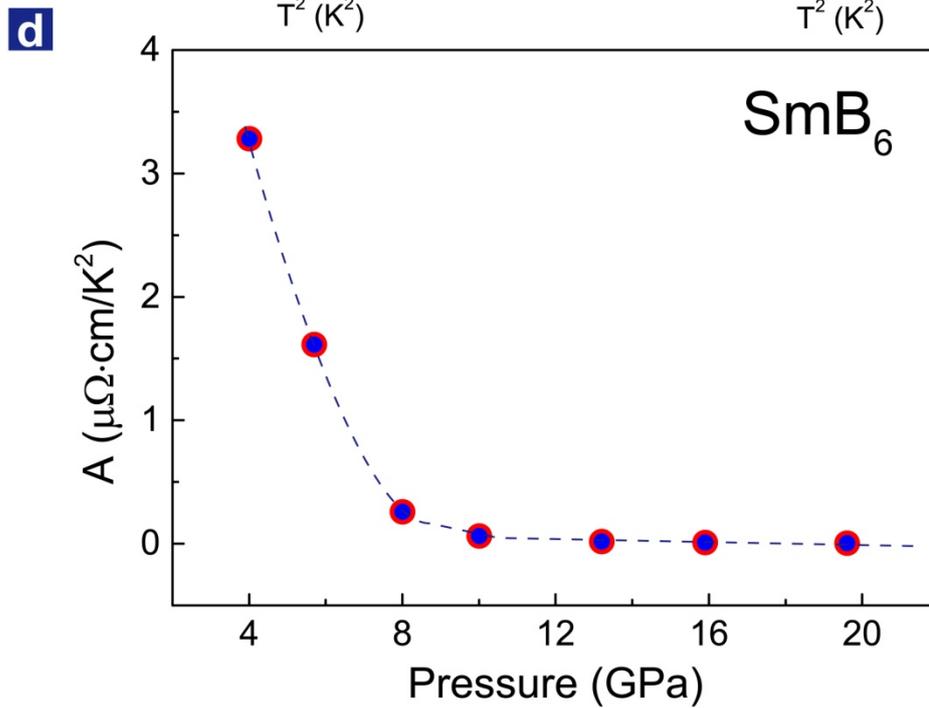

Figure 2 Electrical resistivity as a function of temperature measured down to 0.3 K. (a) The blue solid are the plot of resistivity versus temperature obtained at 4 GPa. The data can be fitted by $\rho=\rho_0+AT^2$ in the range below 0.7 K (red line) and by $\rho=\rho_0+BT$ in the temperature range of 1.5-4 K (light blue), respectively. The low temperature part is zoomed in for a clear view (inset), indicating a Fermi-liquid ground state. (b) and (c) Electrical resistivity versus $T^2$ at 5.7 GPa and 13.2 GPa, respectively. The highest temperature of the quadratic behavior is defined as the onset temperature ($T_{FL}$) of the Fermi-liquid state. (d) Pressure dependence of the $A$ coefficient, showing a divergence tendency as the pressure approaches ~ 4 GPa.

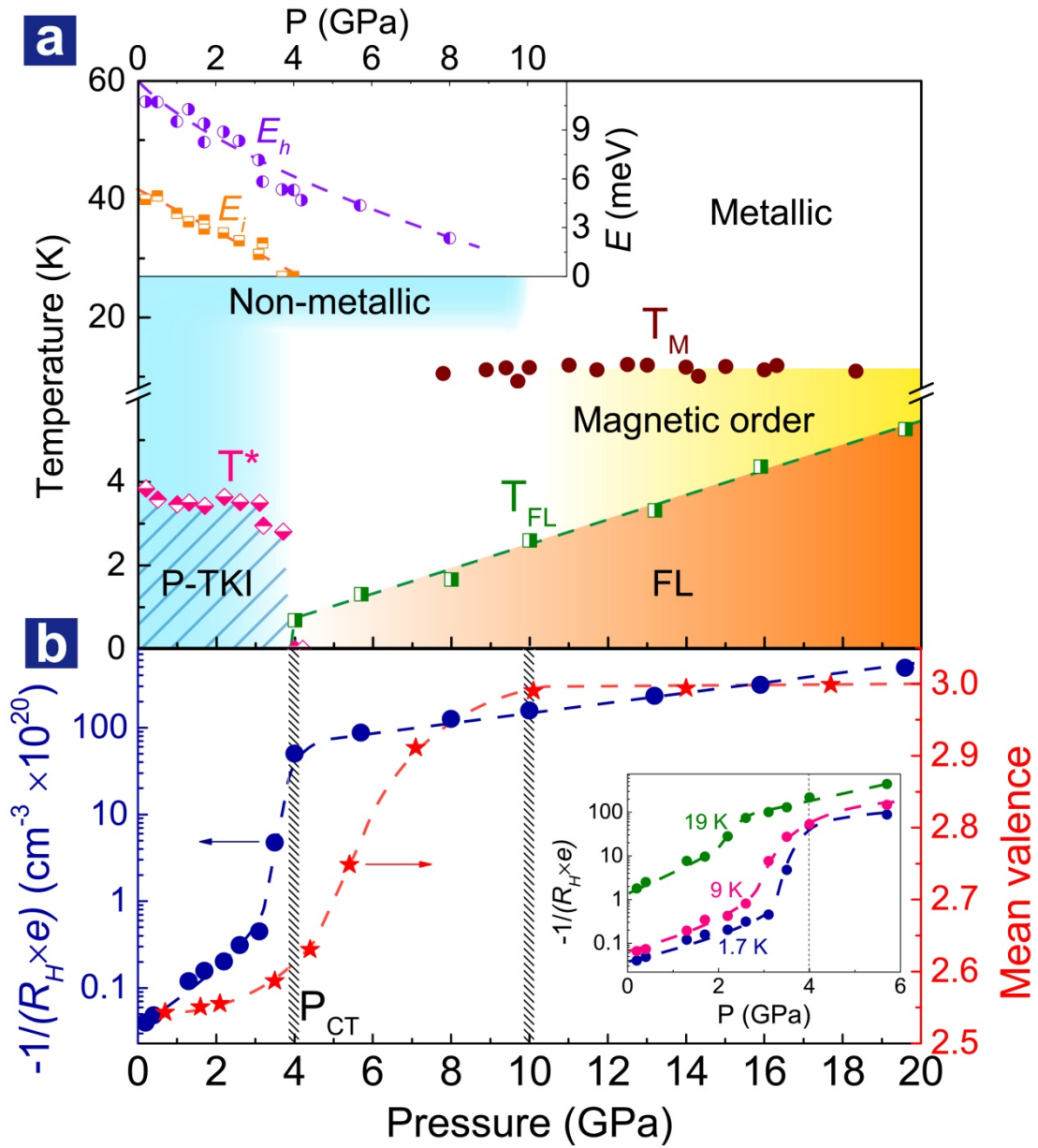

**Figure 3 Summary of pressure dependence of temperature, inverse Hall coefficient and mean valence of SmB$_6$.** (a) Phase diagram depicted by the pressure dependence of the onset temperature of the resistance plateau ($T^*$) and Fermi liquid ($T_{FL}$), respectively. Here, P-TKI and FL represent putative topological Kondo insulating state and conventional Fermi liquid state, respectively. $T_M$ stands for the magnetic transition temperature taken from Ref. [29,30]. The inset displays pressure dependence of the two activation gaps $E_i$ and $E_h$, respectively. (b) Plots of inverse Hall

coefficient ($R_H$) and mean valance of Sm ions versus pressure. The inset displays the inverse $R_H$ as a function of pressure at different temperatures.